\documentstyle[12pt,twoside,fleqn,espcrc1,psfig,times]{article}

\newcommand{\AmS}{{\protect\the\textfont2
  A\kern-.1667em\lower.5ex\hbox{M}\kern-.125emS}}
\title{Strange Antibaryons from QGP}
\author{Jean Letessier\address{Laboratoire de Physique 
Th\'eorique et Hautes Energies$^1$,\\
Universit\'e Paris 7, 2 place Jussieu, F--75251 Cedex 05},
Johann Rafelski\address{Department of Physics, University of Arizona, 
Tucson, AZ 85721}
and Ahmed Tounsi$^{\mathrm a}$}

\begin{document}
\setcounter{footnote}{1}
\footnotetext{Unit\'e  associ\'ee au CNRS UA 280.}
\maketitle
\vspace*{-6cm}
\centerline{Presented at Quark Matter 1995, Monterey, {\it Nucl. Phys.}
{\bf A590} (1995) 613c.}
\vspace*{5.5cm}
\begin{abstract}
We study as function of the collision energy and stopping the thermal
conditions reached in a quark--gluon plasma fireball formed in a 
relativistic heavy ion collision. We explore strange particle 
yields for the current round of Pb--Pb and Au--Au experiments. 
\end{abstract}

\section{INTRODUCTION}
Recently results have been presented which allow a test of the hypothesis
made long ago \cite{predict} that strange particle production could be 
providing key information about the formation and properties of the QGP 
phase in relativistic heavy ion collisions at CERN (160--200 GeV A) 
and/or AGS (10--15 GeV A) energies. A strong enhancement of all strange particle 
yields relative to proton-proton and proton-nucleus collisions has been
well documented --- the more specific strange antibaryon 
enhancement is reported at central rapidity by the WA85/94 \cite{WA85} 
and NA35 \cite{NA35} collaborations in 200 GeV A collisions of sulphur 
nuclei with various nuclear targets. 
 
In the picture of sudden hadronization of the QGP fireball, as well as
when strange particles are made in radiative/recombination formation on
the surface during expansion, the light quark fugacity determines the
particle ratios. If during the final state hadronization a full 
re-equilibration occurs, then the memory of the QGP would be largely lost
--- except that under certain conditions the excess entropy produced in
such events leaves a characteristic trace. However, a phenomenological
analysis \cite{analyze} of data taken at 200 GeV A suggests that we have
a very favorable situation in this case: the particle flow results
suggest strongly that the picture of a rapidly disintegration central
fireball applies, and the emitted particles carry in this case
information about properties of the primordial QGP state.
 
Given this observation, we have here as objective to develop a 
first-principle model that would allow to describe current data. What is
needed is a condition which would express how much nuclear matter can be
compressed in the collision as the constituents thermalize. We elaborate
here on the experimental consequences of the proposal we made before 
\cite{LET9423} that the condition of balance between the thermal pressure
of the deconfined quarks and gluons and the dynamical compression
exerted by the  in-flowing  nuclear matter during the interpenetration
of the projectile and the target is determining the conditions at
particle freeze-out.
 
\section{\uppercase{Initial Fireball Conditions}}
The tacit assumption of rapid thermal equilibration in primary
interactions introduces almost 70\% of the final state entropy into the
initial state. This entropy manifests itself later as the multiplicity of
produced particles. Our approach associates this initial entropy content
with certain initial statistical conditions. A study undertaken
specifically to understand if the initial state assumptions are impacting
the final state \cite{cool} has shown that the crucial parameter is, for
a thermally isolated system, the fireball energy per baryon. Using
applicable equations of state the dependence of $E/B$ on the statistical
parameters is determined. On the other hand, introducing the stopping
fractions of energy $\eta_{\rm E}$ and baryon number $\eta_{\rm B}$
we obtain: \begin{equation} \label{ECM}
{E\over B}\equiv {\eta_{\rm E}{E_{\rm CM}}\over {\eta_{\rm B}A_{\rm
part}}} \simeq {E_{\rm CM}\over A_{\rm part}}\,,
\end{equation}
where the last equality assumes that both stopping fractions are equal 
--- this we believe to be a reasonable assumption for the current experimental
domain. $A_{\rm part}$ is the number of nucleons participating in the
reaction. In consequence, the energy per baryon in the fireball is to be
taken as being equal to the kinematic energy available in the collision.
In the present day experiments we have the following conditions: for 
Au--Au at AGS --- $E/B =2.6$ GeV, for S--W/Pb at CERN ---$ E/B=8.8$ 
GeV and for Pb--Pb  at CERN --- $ E/B=8.6$ GeV. 
 
After the initial thermalization occurred, the internal pressure of the
fireball will be determined considering balance between the thermal
pressure of a quark-gluon plasma and the dynamical compression exerted
by the in-flowing matter, augmented by the external vacuum pressure.  We
employ perturbative QCD, QGP-equation of state with thermal masses.
Aside of the temperature $T$ we encounter the different (quark and gluon)
fugacities $\lambda_i, i=q,s,G$ and the chemical saturation factors
$\gamma_i$ for each particle. For the vacuum pressure we will use
${\cal B}\simeq0.1 \mbox{GeV/fm}^3$. The pressure due to kinetic motion
follows from well-established principles, and we obtain:
\begin{equation}\label{Pbal}
P_{\rm th}(T,\lambda_i,\gamma_i)=P_{\rm dyn}+P_{\rm vac}
=\eta_{\rm p} \rho_{0}\frac{p_{\rm CM}^2}{E_{\rm CM}}+{\cal B}
\end{equation}
Here $\rho_{0}$  is the normal nuclear density and it is understood
that the energy $E_{\rm CM}$  and the momentum $p_{\rm CM}$ are given in
the nucleon--nucleon CM frame  and $\eta_{\rm p}$ is the momentum
stopping fraction  --- only this fraction $0\le\eta_{\rm p}\le 1$ of the
incident CM  momentum can be used  by a particle incident on the central 
fireball  (the balance remains in the un-stopped longitudinal motion) in
order to exercise dynamical pressure. For a target transparent to the
incoming flow, there would obviously be no pressure exerted. 
 
Given values of $E_{\rm CM}$ and $\eta_{\rm p}$, as well as taking
$\eta_{\rm E}=\eta_{\rm B}$ we have prescribed a particular energy per
baryon in the fireball and we can concurrently solve the pressure
equilibrium constraint Eq.\,(\ref{Pbal}). These two conditions allow to
determine two initial statistical parameters. We choose these to be the
temperature $T_{\rm th}$ and the initial value of quark fugacity
$\lambda_{\rm q}$. Because the QGP phase is strangeness neutral we have
$\lambda_{\rm s}=1$.
\begin{figure}[t]
\vspace*{-5.5cm}
\centerline{\hspace*{1.7cm}
\psfig{height=15cm,figure=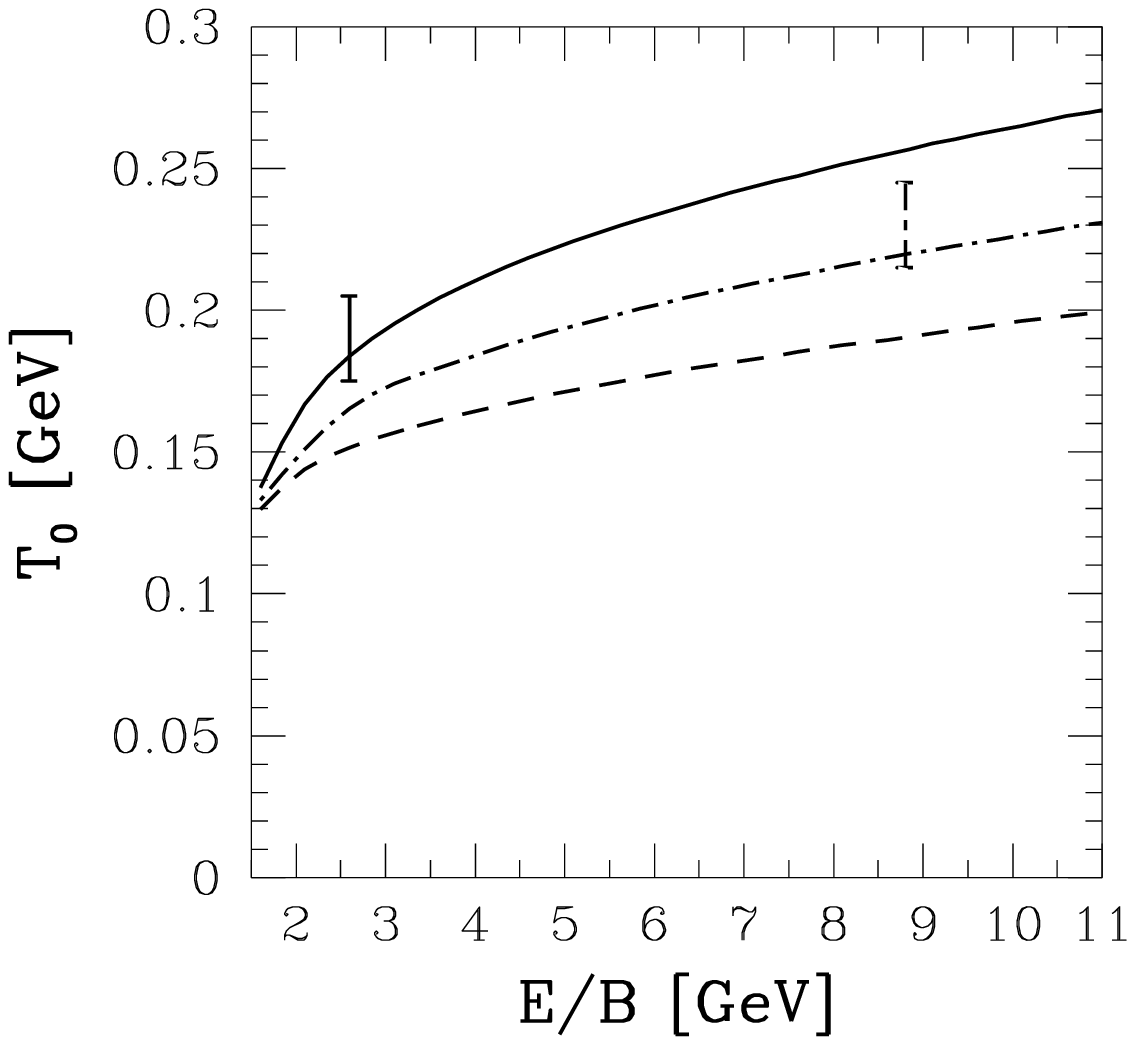}
\hspace*{-2.6cm}\psfig{height=15cm,figure=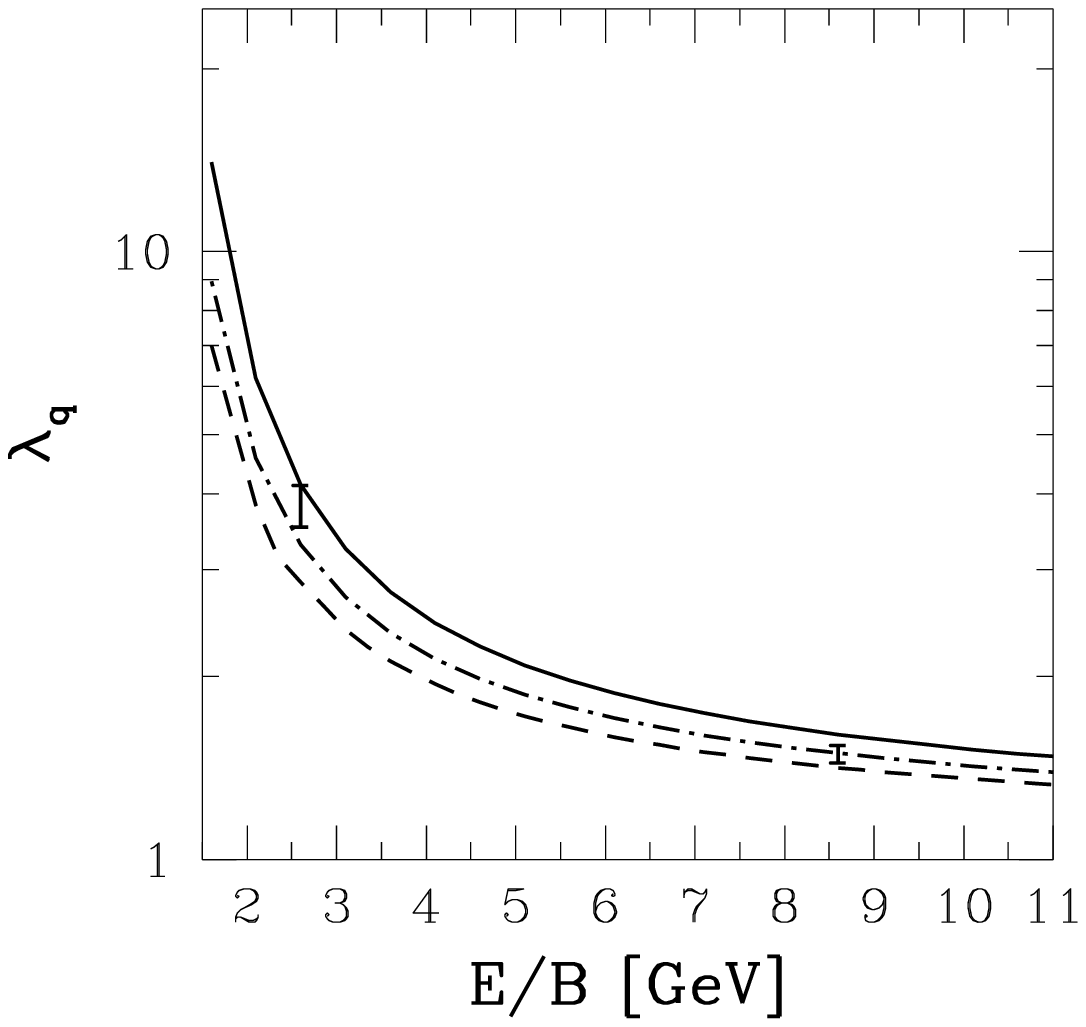}} 
\vspace*{-3.2cm} 
\caption{\protect\small 
Temperature $T_0$  and light quark fugacity $\lambda_{\rm q}$
at the time of full chemical equilibration as function of the
fireball energy content $E/B$. The range of the possible values as
function of $\eta$ is indicated by showing results, for
$\eta=1$ (solid line), 0.5 (dot-dashed line) and 0.25 (dashed line). 
\protect\label{fig1QM95}
}
\vspace*{-.5cm}
\end{figure}
 
During the time the nuclei interact, we evolve the fireball
statistical parameters at constant pressure and energy per baryon. 
We need not to establish a time scale for the change in time of 
both $\gamma_{\rm q}$ and $\gamma_{\rm G}$ as we assume that it
occurs at equal rate. We also take that the strange quark chemical 
relaxation constant is about 7 times smaller than the one for light 
quarks ---  hence while the values of the chemical
occupancy factors for light quarks ($\gamma_{\rm q}\to 1$) and gluons 
($\gamma_{\rm G}\to 1$) approach chemical equilibrium, for the strange 
quarks $\gamma_{\rm s}\to 0.15$. In the final evolution step, which 
occurs in our approach after the collision has ended, thus for proper 
times $t\ge1$fm/c, but probably also at $t\le$ 3--5 fm/c, we relax the 
strange quarks (near) to their equilibrium abundance, and the temperature
drops to the value $T_0$ --- we presume that while this last stage of
the temporal evolution occurs, the entropy of the light quarks and gluons 
remains nearly constant.
 
In Fig.\,\ref{fig1QM95} we show the behavior of temperature $T_0$  and
the light quark fugacity $\lambda_{\rm q}$ at the time of full  chemical
equilibration, as function of the fireball energy content $E/B$ (energy
per nucleon in the CM frame). It is believed that in current 200 GeV 
experiments $\eta\equiv\eta_{\rm E}\simeq\eta_{\rm B}\simeq\eta_{\rm p}
\simeq 0.5$ while at the lower AGS energies $\eta\simeq 1$. The range of
the possible values as function of $\eta$ is indicated by showing in 
Fig.\,\ref{fig1QM95} results for $\eta=1$ (solid line), 0.5 (dot-dashed 
line) and 0.25 (dashed line). These results 
are in quantitative agreement with present day experiments as is indicated 
by the vertical error lines drawn in Fig.\,\ref{fig1QM95}. 
Our simple model thus spans correctly the large variation 
domain of the statistical fireball parameters as function 
of collision energy and stopping power.
\nopagebreak 
\section{\uppercase{Strange particle ratios}}
\begin{figure}[t]
\vspace*{-1.8cm}
\centerline{\hspace*{-2.9cm}
\psfig{height=18.2cm,figure=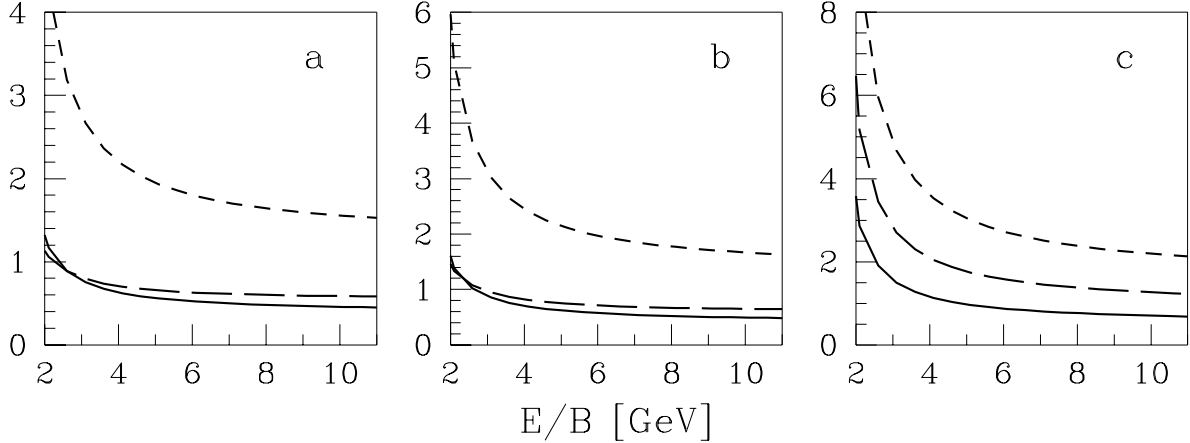}
\hspace*{-3.5cm}}
\vspace*{-11.5cm}
\caption{\protect\small 
Strange antibaryon ratios $\overline{\Lambda}/\overline{p}$ (short-dashed line),
$\overline{\Xi^-}/\overline{\Lambda}$ (solid line) and
$\overline{\Omega}/\overline{\Xi^-}$ (long-dashed line) are displayed as
function of $E/B$; in {\bf a} for the entire range of $p_\bot$, in {\bf
b} for particles with $p_\bot\ge 1$ GeV and in {\bf c} for 
particles with $m_\bot \ge 1.7$ GeV.
 \protect\label{fig2QM95}}
\vspace*{-.5cm}
\end{figure}

Assuming direct hadronization of QGP, we compute the resulting particle
ratios for $\eta=1$, applicable to the collisions of heaviest nuclei. 
For the strange antibaryons the three most interesting 
ratios are: $\overline{\Lambda}/\overline{p},\ \overline{\Xi^-
}/\overline{\Lambda}, \ \overline{\Omega}/\overline{\Xi^-}\,. $
These are displayed in the Fig.\,\ref{fig2QM95}; in \ref{fig2QM95}{\bf a}
for the entire range of $p_\bot$, in \ref{fig2QM95}{\bf b} for particles
with $p_\bot\ge 1$ GeV and in \ref{fig2QM95}{\bf c} for particles
with $m_\bot \ge 1.7$ GeV. We encounter in all cases the remarkable
behavior that all these three ratios increase as the collision energy is
reduced (for not full chemical equilibrium, multiply each ratio with the
value of $\gamma_{\rm s}$).
We further note that with rising strangeness content the ratios shown in
Fig.\,\ref{fig2QM95} get bigger, the ratios at fixed $m_\bot$ are all
greater than unity. We have also computed the absolute yields of
these particles. They drop rapidly with energy --- at AGS the yield is
typically a factor 100 smaller compared to SPS energies.
 
We have developed a simple dynamical model allowing to determine in a
systematic fashion the conditions reached in high density matter
generated in heavy ion collisions. We analyzed here the situation
assuming the QGP equation of state of dense matter. Comparison of our
results for a reasonable choice of residual and kinematic parameters with
the observed properties in \mbox{S--W/Pb} and Si--Au collisions shows a high
degree of quantitative agreement between our QGP-fireball model and the
experimental results. 
\subsection*{Acknowledgments} 
\hspace*{\parindent}  
J. Rafelski acknowledges partial support by US DOE grant
DE-FG02-92ER40733.
 

\end{document}